\documentstyle[aps, prd,epsfig,subeqnarray]{revtex}
\draft 
\begin{document}
\def\ok{\stackrel{0}{K}}

\twocolumn[\hsize\textwidth\columnwidth\hsize\csname
@twocolumnfalse\endcsname
%

%
%
\title{Initial Data and Coordinates for Multiple Black Hole 
Systems} 
\author{Richard A. Matzner$^{\ast \dagger}$ \\
Mijan F. Huq$^{\ast}$ \\
Deirdre Shoemaker$^{\ast}$}
\address{ $^{\ast}$ Center for Relativity, The University of Texas at Austin, 
Austin, TX 78712-1081 \\ 
$^{\dagger}$Orson Anderson Scholar, Los Alamos National Laboratory, 1996-97 }

\maketitle
\begin{abstract}
We present here an alternative approach to data setting for spacetimes 
with multiple moving black holes generalizing the Kerr-Schild 
form for rotating or non-rotating single black holes to multiple moving 
holes. Because this scheme preserves the Kerr-Schild form near the holes, 
it selects out the behaviour of null rays near the holes, may simplify 
horizon tracking, and may prove useful in computational applications.  
For computational 
\textit{evolution}, a discussion of coordinates (lapse function and shift 
vector) is given which preserves some of the properties of the single-hole 
Kerr-Schild form. 
\end{abstract}

\vskip2pc]

\section{Introduction}
In the numerical simulation of gravitational spacetimes Einstein's equations
are cast as a Cauchy problem in the ADM 3+1 formalism. In this form the 
equations are split into a set of evolution and constraint equations. The
latter set, the Hamiltonian and momentum constraints, are elliptic
partial differential equations which must be imposed at the initial slice in
the Cauchy evolution. Thereafter, analytically, the evolution equations
preserve them. York, in a series of papers\cite{17,YOM,B+York,YorkMS} 
gave a framework for solving the constraint equations known as the
conformal formalism. We take $G=1$ (Newton's constant) and $c=1$ (the speed of
light).  The 3-metric, $g_{ij}$, is assumed to be of the form
\begin{eqnarray}
g_{ij} & = & \phi^4 \hat{g}_{ij}
\label{3metric.1}
\end{eqnarray}
where $\hat{g}_{ij}$ is the base metric; $g_{ij}$ is then termed the
physical metric and $\phi$ is the conformal factor. The extrinsic curvature
is decomposed into its trace and trace-free parts:
\begin{eqnarray}
\hat{K}_{ij} & = & \hat{E}_{ij} + \frac{1}{3} \hat{g}_{ij} K
\label{extcurv.1}
\end{eqnarray}
where again the hatted quantities refer to the tensors in the base metric
and $\hat{E}_{ij}$ is the traceless part of the extrinsic curvature. 
In this approach $K$ is assumed to be a given scalar function.
The physical extrinsic curvature is related to the base extrinsic 
curvature through the following conformal transformation:
\begin{eqnarray}
E_{ij} & = & \phi^{-10} \hat{E_{ij}}.
\label{extcurv.2}
\end{eqnarray}
The Hamiltonian and momentum constraints then can be written as:
\begin{eqnarray}
8 \hat{\Delta} \phi - \hat{R} \phi - \frac{2}{3} K^2 \phi^5 + 
E^{ij} E_{ij} \phi^{-7} & = & 0
\label{hamconst.1}
\end{eqnarray}
and
\begin{eqnarray}
\hat{D}_j E^{ij} - \frac{2}{3} \hat{D}^i K & = & 0
\end{eqnarray}
where $\hat{\Delta}$ is the laplacian with respect to $\hat{g}_{ij}$,
$\hat{D}$ is the covariant derivative compatible with $\hat{g}_{ij}$
and since we have a vacuum spacetime we take the matter terms to be zero.

York's method, as it has been applied in the  black hole case, assumes that 
the 3-space is conformally
flat, with holes, and that the expansion of the 3-space $(Tr K)$ vanishes.  
For the case we consider, we assume the initial spatial domain contains holes.
A method has been given\cite{kulkarni.etal,Cook91} for specifying 
an essentially analytic solution to 
the momentum constraint with symmetric boundary conditions (an infinite series 
has to be summed) for a multiple black hole spacetime.  The remaining 
difficulty is then to solve the Hamiltonian 
constraint (\ref{hamconst.1}), which is a nonlinear elliptic equation, for the 
conformal factor.  A 
completely convergent method for doing this was exhibited\cite{18}, where a 
controllable convergent algorithm produces solutions of specifiable 
accuracy.
The static single hole initial-data solution found in this manner is a 
$t=$constant slice of the isotropic coordinate representation of 
Schwarzschild\cite{Schw}. (One of the features is that the initial slice 
does not penetrate \textit{inside} the horizon.)  Multiple moving spinning 
black holes can be specified. 

A new alternative to the conventional 
method based on throats and conformal imaging was proposed and implemented
by Brandt and Br\"{u}gmann\cite{brandt} where the black holes are treated as 
punctures. The internal asymptotically flat regions are compactified to
obtain a domain without inner boundaries leading to significant 
computational simplications. 

Recent computational black hole work has turned to solutions of the 
Kerr-Schild\cite{16,Kerr63} form, because these analytical models 
have no coordinate pathology at the horizon and allow slicings which penetrate 
within the 
horizon\cite{choptuik.1}. The present philosophy of handling the singularity 
hidden within 
a black hole is to compute only up to the horizon
\cite{thornburg,seidel.suen}, though such computational 
techniques almost certainly require a consideration of points slightly within 
the horizon. Further, having an analytic, tractable example is a 
great asset in developing a computational scheme. Hence we propose a scheme
where we utilize our freedom in setting data for the base 3-metric and 
extrinsic curvature and use this analytical specification based on the
Kerr-Schild form.

The motivation for this work is to initialize computational evolutions being
done by the Binary Black Hole Grand Challenge Alliance\cite{bbh} which has
shown some success at evolutions based on the Kerr-Schild form.
The approach here focuses on the 3-dimensional initial value problem,
and puts less emphasis on  a representation which resembles the Kerr-Schild
one in a 4-dimensional sense. A complementary approach \cite{bishop.1} within
the Alliance, which
does emphasize the 4-dimensional aspects has recently appeared. The two 
approaches differ significantly, as we discuss more completely in section VII
below.

\section{Kerr-Schild forms for Isolated Black Holes} 

The Kerr-Schild spacetime metric is given by:
\begin{equation}
	ds^{2} = \eta_{\mu \nu} dx^{\mu} dx^{\nu} + 2H(x^{\alpha}) l_{\mu} 
	l_{\nu} dx^{\mu} dx^{\nu}
	\label{1}
\end{equation}
where $\eta_{\mu \nu}$ is the usual flat space form, $H$ is a scalar 
function of position and time and $l_{\mu}$ is an (ingoing) null vector (null 
both in the background, and in the full metric),
\begin{eqnarray}
& & ~~~~~~~~~~
\eta^{\mu \nu} l_{\mu} l_{\nu} = g^{\mu \nu} l_{\mu} l_{\nu} = 0,
\label{2}
\end{eqnarray}
so that $l_t^2 = l_i l_i$.
It is the fact that the Kerr-Schild form is based on and selects null 
surfaces near the 
black hole which makes it so appealing as a scheme for setting coordinates.
The general Kerr-Schild black hole metric (written in Kerr's original 
rectangular coordinates) has
\begin{equation}
	H = \frac{Mr}{r^{2} + a^{2}\cos^{2} \theta}
	\label{3}
\end{equation}
and
\begin{equation}
	l_{\mu} = \left(1, \frac{rx + ay}{r^{2} + a^{2}}, \frac{ry - 
	ax}{r^{2} + a^{2}}, \frac{z}{r}\right). 
	\label{4}
\end{equation}
Here $M$ is the mass of the Kerr black hole and $a = J/M$ is the 
specific angular momentum of the black hole, $r,~ \theta$ (and $\phi$) 
are auxiliary spheroidal coordinates.  $z = r \cos \theta$ and $\phi$ is 
the axial angle.  $r(x, y, z)$ is obtained from the relation,
\begin{equation}
	\frac{x^{2} + y^{2}}{r^{2} + a^{2}} + \frac{z^{2}}{r^{2}} = 1 
	\label{5}
\end{equation}
as
\begin{equation}
	r^{2} = \frac{1}{2}(\rho^{2} - a^{2}) +
	\sqrt{\frac{1}{4}(\rho^{2} - a^{2}) + a^{2}z^{2}}, 
	\label{6}
\end{equation}
with $\rho = \sqrt{x^{2} + y^{2} + z^{2}}$. Notice that 
the function $H$ given by Eq.~(\ref{3}) is harmonic:
\begin{equation}
	\nabla^{2}H = 0
	\label{7}
\end{equation}
where this $\nabla^{2}$ is the flat space background Laplacian.
In the limit $r \rightarrow \infty$ or $a \rightarrow 0$ 
we recover the Schwarzschild metric in Kerr-Schild form 
$(H = M/r, ~ l_{\mu} = (1, n_{i}))$.  This corresponds to the ingoing
Eddington-Finkelstein \cite{E-F}
 form of the Schwarzschild metric.
Notice that for the stationary Kerr 
(and Schwarzschild) black holes we choose a particular normalization of 
$l_{\mu}$:$l_{t} \equiv l_{\mu} t^{\mu} = 1$.
In the 3+1 form used in many computational approaches, one splits the 
4-metric into a 3-metric, a lapse and a shift.
The 3-metric gives distances measured in a given spacelike 3-surface (at a
particular instant of time). The inverse of the lapse function gives the
ratio of the coordinate time (measured normally to the $t=$ const 3-space) to
the evolution of proper time in moving along that timelike direction.
The shift vector (multiplied by the interval of coordinate time $dt$ ) gives
the amount that coordinate labels shift in going from one $t=$constant 
3-space to the next.

The general Kerr-Schild metric can be cast in a 3+1 form: 
\begin{eqnarray}
	(\mbox{lapse}) ~ \alpha 
	& = & \frac{1}{\sqrt{1 + 2 Hl_{t}\,^{2}}}
	\label{8}  \\
	(\mbox{shift}) ~ \beta_{i} 
	& = & 2Hl_{t}l_{i}
	\label{9}  \\
	(\mbox{3-metric}) ~ g_{ij} 
	& = & \delta_{ij} + 2Hl_{i}l_{j}.
	\label{10}
\end{eqnarray}
For black hole spacetimes the relation between the lapse and the shift obtained
from this splitting
guarantees that the horizon stays at a constant location, even though
the lapse is non-zero at the horizon. This is in contrast to the isotropic
coordinate representation of the Schwarzschild solution, in which the shift
vanishes everywhere and the lapse vanishes at the horizon.
Hence, for example, the static Schwarzschild black hole (the 
``Eddington-Finkelstein\cite{E-F} form'') : 
\begin{eqnarray}
	&  & H = M/r
	\label{11}  \\
	\alpha^{2} & = & \frac{1}{1 + 2M/r}\,,
	~~ \beta_{i} = \frac{2M}{r} \frac{x_{i}}{r} 
	\label{12}  \\
	&  & g_{ij} = \eta_{ij} 
	+ \frac{2M}{r} \frac{x_{i}}{r} \frac{x_{j}}{r}\,,
	\label{13} \\
K_{ij} &=& \frac{2M}{r^4} \frac{1}{\sqrt{1+2M/r}} \left[ r^2 \eta_{ij} - (2 + \frac{M}{r}) x_i x_j \right],
\end{eqnarray}
is smooth at the horizon $r = 2M$.  Notice that the 
nonmoving Eddington-Finkelstein metric uses 
\begin{equation}
	l_{i} = \partial_{i}(M/H),
	\label{14}
\end{equation}
where $\partial_i$ is the partial derivative.

\section{Boosted Black Holes} 

The Kerr-Schild metric is form-invariant under a boost due to its structure.
This makes it ideal as a metric for constructing a model problem for moving
black holes\cite{Huq}.  One simply applies a 
constant Lorentz (boost velocity ${\bf v}$ as specified in the background
Minkowski spacetime) transformation $\Lambda^{\alpha}_{\beta}$ to the 
4-metric. The resulting metric retains the Kerr-Schild form, but with 
straightforwardly transformed $H$ and $l_{\mu}$:
\begin{equation}
	\begin{array}{rcl}
		x'^{\beta} & = & \Lambda_{\alpha}^{\beta} x^{\alpha}  
		 \\
		H(x^{\alpha}) 
		& \rightarrow &  H(\Lambda^{-1 \alpha}\,_{\beta} x'^{\beta}) 
		\\
		l'_{\delta} 
		& = & \Lambda^{\gamma}\,_{\delta} 
		l_{\gamma}(\Lambda^{-1 \, \alpha}\,_{\beta} x'^{\beta}) 
		 \\
		g'_{\mu \nu} 
		& = & \eta_{\mu \nu} + 2H l'_{\mu} l'_{\nu}
	\end{array}
	\label{15}
\end{equation}
Under this boost, $l_{t}$ is no longer unity. Note that 
because we start with a stationary solution the only 
time dependence is in the motion of the center.  In this sense the 
solution has no ``extra'' radiation loaded into the initial data for 
the boosted hole (since we see none escape to infinity).

Under such a boost, $H$ no longer satisfies Eq.~(\ref{7}), but every 
solution of Eq.~(\ref{7}) in the nonmoving frame satisfies
\begin{equation}
	\nabla^{2}H - ({\mathbf{v}} \cdot \mathbf{\nabla})^{2} H = 0
	\label{16}
\end{equation}
in the boosted frame, where $\mathbf{\nabla}$ is the flat background 
spatial derivative
operator.

\section{Boosted Schwarzschild} 

For example, for the Eddington-Finkelstein system boosted in the 
$z$-direction ($v \equiv v^z$), with new coordinates $(x, y, z, t)$ we have
\begin{equation}
	\begin{array}{lcl}
		r^{2} & = & x^{2} + y^{2} + \gamma^{2}(z - v t)^{2} \\
		l_{t} & = & \gamma(1 - v \gamma (z - v t)/r) \\ 
		l_{x} & = & x/r \\
		l_{y} & = & y/r \\
		l_{z} & = & \gamma(\gamma (z - v t)/r - v) \\
		 H & = & M/r  \\
	\end{array}
	\label{17}
\end{equation}
where $\gamma = 1/\sqrt{(1-v^2)}$

Under a boost, the metric becomes explicitly time dependent (because 
$r$ is time dependent).

Because the boost of the Schwarzschild solution merely ``tilts the 
time axis,'' we can consider all of the boosted 3+1 properties at an instant 
$t = 0$, in the frame which sees the hole moving.  Subsequent times 
$t$ simply offset the solution by an amount $vt$.

After the boost, $l_{i}$ no longer solves Eq.~(\ref{14}), but
\begin{equation}
	l_{i} = \partial_{i}(M/H) - \gamma v_{i}\,.
	~~\mbox{(Boosted E-F)}
	\label{18}
\end{equation}
With Eq.~(\ref{16}) and Eq.~(\ref{18}) $\alpha, ~ \beta^i$ are defined via 
Eq.~(\ref{8}) and Eq.~(\ref{9}).  Figure (1) shows a slice through an 
Eddington-Finkelstein solution boosted to $0.5c$ (moving upward in the figure).
The heavy inner contour is 
the horizon, the cardioid contours are lines of constant $\alpha$, and the 
line segments indicate the direction and magnitude of $\beta_i$.

The extrinsic curvature for the boosted Kerr-Schild metric is given 
by computing $\dot{g}_{ij}$ in the boosted frame (given by combining 
the boosted terms from Eq. (\ref{15}) into Eqs. (\ref{8})-(\ref{10})).

Now, one of the Einstein equations is
\begin{equation}
	\dot{g}_{ij} = \beta_{i, j} + \beta_{j, i} - 2 \Gamma^{k}_{ij} 
	\beta_{k} - 2 \alpha K_{ij}\,,
	\label{19}
\end{equation}
where the dot indicates partial time derivatives.
Eq.~(\ref{19}) enables the computation of $K_{ij}$.

We find that
\begin{equation}
	K_{ij} \sim O(\frac{M}{r^{2}})(1 + O(v)).
	\label{21}
\end{equation}

\begin{figure}
 	\centering
  \epsfxsize=7.5cm
  \begin{center}
   \leavevmode
  \epsffile{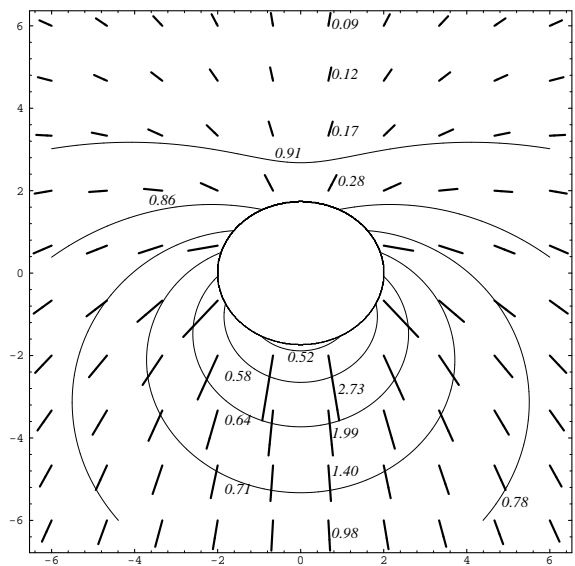}
   \end{center}
	\caption{Contours of lapse and shift vector $\beta_{i}$ for a boosted 
	Schwarzschild (Eddington-Finkelstein) black hole.  This is a cut 
	through the equator; the hole is boosted at 0.5 upward in the 
	figure.  The shift is smaller ahead of the hole than behind, which 
	allows the black hole to move through the computational grid.  The 
	ovoid figure is the horizon (distorted in these coordinates).  The 
	cardioid curves are contours of constant lapse $\alpha$, with values from 
	the topmost contour down of 0.91, 0.86, 0.78, 0.71, 0.64, 0.58, 
	0.52.  Along the axis of motion the lapse for this $v=0.5$ case is
   $\sqrt{3}/2$ at the leading point and $1/2$ at the trailing point.}
	\label{Figure 1}
\end{figure}

\section{Setting Multiple Black Hole Data} 

We will work by setting data for \textit{two} Kerr-Schild black 
holes of comparable mass $M_{1} \sim M_{2}$. For purposes of 
development here (e.g. to estimate the size of terms), we also 
assume $M_{1}/r_{12} \ll 1$ where $r_{12}$ is the coordinate 
separation between holes. Each hole has a velocity 
${\mathbf{v}}_{1}$ or ${\mathbf{v}}_{2}$, as appropriate, assigned to it. 

Define
\begin{eqnarray}
r_{1}\,^{2} & = & (x - x_{1})^{i} (x - x_{1})^{j} \delta_{ij}\,, 
\nonumber \\ 
r_{2}\,^{2} & = & (x - x_{2})^{i} (x - x_{2})^{j} \delta_{ij}
\label{22} 
\end{eqnarray}
where $x_{1}^{i}$ and $x_{2}^{i}$ are the coordinate positions of the holes on 
the initial slice.

Here we use our freedom to set data and fix the background 3-metric as follows:
\begin{eqnarray}
	& ^{(3)}\hat{d}s^{2} = \hat{g}_{ij} dx^{i} dx^{j} 
	= \delta_{ij} dx^{i} dx^{j} 
	+ {}_{1}\!\!H(r_{1})_{1}l_{i1}l_{j}dx^{i}dx^{j}  & 
	\nonumber \\
	& + {}_{2}H(r_{2})_{2}l_{i2}l_{j}dx^{i}dx^{j}  & 
	\label{23}
\end{eqnarray}
Here $_{a}H$, and $_{a}l_{i} (a = 1, 2)$ are the functions defined from 
each single (perhaps boosted) black hole.  This background metric has 
two vectors corresponding to the null vector of the Kerr-Schild form 
(although at this point we see only their spatial components). 

The $~ \hat{} ~$ symbol indicates that this is a conformally related 
metric, while the physical metric is
\begin{equation}
	g_{ij} = \phi^{4} \hat{g}_{ij}
	\label{24}
\end{equation}
Here $\phi$ is a strictly positive conformal factor which will be 
determined in the process of solving the constraints. 

We start the constraint-solution process with a trial 
$\hat{K}_{a}\,^{b}$:
\begin{equation}
	_{0}\hat{K}_{a}\,^{b} = \hat{K}_{a}\,^{b}(1) + \hat{K}_{a}\,^{b}(2) 
	\label{25}
\end{equation}
These $\hat{K}_{a}\,^{b}(1), \hat{K}_{a}\,^{b}(2)$ are the individual 
extrinsic curvatures, computed and indices raised using the 
single-hole boosted Kerr-Schild metric appropriate to either $M_{1}$ 
or $M_{2}$. (Henceforth we use the 2-hole physical or conformal 
metric.) The leading subscript on the left indicates that this is a 
zeroth order approximation (in the sense of ``zeroth'' guess; this 
is \textit{not} an iteration method).

Following York, we separate the trace: $K = \hat{K}(1) + \hat{K}(2)$ 
from the traceless part of $_{0}\hat{K}_{ab}$: 
\begin{equation}
	_{0}\hat{E}^a_b = ~_{0}\!\hat{K}^a_b - \frac{1}{3} 
	\delta^a_b K.
	\label{26}
\end{equation}
$K$ is considered a given scalar function and is not conformally 
scaled. The conformal scaling for $_{0}\hat{E}{^{ab}}$ 
is chosen as:
\begin{equation}
	_{0}\!E{^{ab}} = 
	\phi^{-10}\,_{0}\hat{E}{^{ab}}\,.	
	\label{27}
\end{equation}
Here $_{0}\!E{^{ab}}$ is the traceless part of the extrinsic 
curvature in the physical space associated with the zeroth guess.

We attempt to write the momentum constraint:
 \begin{equation}
	 D_b {}_{0}\!E_{c}{^{b}} - \frac{2}{3}D_c K \not= 0,
	 \label{28}
 \end{equation}
where $D_b$ is the covariant derivative compatible with the 3-metric.
The momentum constraint is violated because of the appearance of 
connections from hole (1) multiplying an extrinsic curvature computed 
from hole (2), and vice versa.

We can solve the momentum constraint equation by adding a term which 
contributes to the longitudinal part of the solution: 
\begin{equation}
	A^{cb} \equiv {}_{0}\!E^{cb} + (lw)^{cb}\,, 
	\label{29}
\end{equation}
where $w^{a}$ is a vector to be solved for\cite{YorkMS}, and 
\begin{equation}
	(lw)^{cb} = D^{c}w^{b} + D^{b}w^{c} 
	- \frac{1}{3}g^{bc}D_{d}w^{d}.
	\label{30}
\end{equation}
We then demand
\begin{subeqnarray}
D_b A^{cb} & - & \frac{2}{3} D^c K = 0, \slabel{31a} \\
~~\mbox{or} & &  \nonumber \\
D_b (lw)^{cb} & = & \frac{2}{3} D^c K - ~
D_b {}_{0}\!E{^{cb}}\,.
\slabel{31b}
\end{subeqnarray}
This is an elliptic equation for $w^{a}$, which gives an addition to 
the extrinsic curvature guaranteeing the solution of the momentum 
constraint. As it stands, however, we cannot directly solve Eq.~(29) 
because it involves the full physical metric, which we have not yet 
specified.

Using the result due to York\cite{YorkMS}:
\begin{equation}
	D_{b} A^{cb} = \phi^{-10} \hat{D}_{b} \hat{A}^{cb}\,, 
	\label{32}
\end{equation}
the momentum constraint can be written as: 
\begin{equation}
	\hat{D}_{b}(\hat{l}w)^{cb} = \frac{2}{3} \hat{g}^{cb} \phi^{6} 
	\hat{D}_{b} K - \hat{D}_{b}(_{0}\hat{E}{^{cb}}) 
	\label{33}
\end{equation}
Here, note
$\hat{D}_{b}K ~\mbox{is just}~ \partial K/\partial x^{b}$. 

Form (\ref{33}) allows solution for $w^{a}$ in the conformal frame, except 
that the value of $\phi$ is not known; in fact this appearance of 
$\phi$ leads to a coupling between this momentum constraint, and the 
Hamiltonian constraint to which we now turn. 

The Hamiltonian constraint solution follows York's development 
exactly. Since $R = \hat{R} \phi^{-4} - 8 \phi^{-5} 
\hat{\bigtriangleup} \phi$ we have
\begin{eqnarray}
	0 = & &  8 \bigtriangleup \phi - \hat{R}\phi - \frac{2}{3} K^{2}\phi^{5} + 
	\phi^{-7} (_{0}\hat{E}{^{ij}} + (\hat{l}w)^{ij}) \nonumber \\ 
	& & (_{0}\hat{E}_{ij} + (\hat{l}w)_{ij}) 
	\label{34}
\end{eqnarray}
Solution of the coupled set Eq.~(\ref{33}) and Eq.~(\ref{34}) for $\phi$ and 
$w$ constitutes a solution of the constraint equations. For boundary conditions 
on $\phi$ we impose $\phi = 1$ at infinity, perhaps through a Robin condition
\cite{York.grav,Cook91}
, and at the surface horizons of the black holes (located in 
the background metric) we set $\phi = 1$.
We impose the condition $w^{a} \rightarrow 0$ at the surfaces of the 
black holes and at $r \rightarrow \infty$. 

First note that if we apply these conditions for a single black hole, since
those exact solutions already satisfy the constraints we obtain a solution
$\phi=1$ and $w^a=0$ immediately.
Then note that if the initial configuration has holes widely separated, we 
expect $\phi \sim 1$ everywhere. Then the momentum equation is an elliptic 
operation for $w^{a}$ with source $\sim r^{-4}$ at large distances 
(since it arises from the Christoffel symbol $\times$ extrinsic 
curvature cross terms). Hence the ``total charge'' of the source is 
well localized, and $w^{a} \sim r^{-1}$, leading to corrections to $K_{ab}$, 
which are the same order in $r$ as the background $K_{ab} \sim r^{-2}$. 
Further, for well-separated holes these corrections are small. In the 
Hamiltonian constraint (\ref{34}), the extrinsic curvature terms are all 
squared, i.e. $\sim r^{-4}$. Hence (again assuming $\phi \sim 1$) the 
elliptic equation for $\phi$ has finite inhomogeneous source $\sim 
K_{ij}\,^{2} \sim r^{-4}$. The remaining question is the behavior of the 
$\hat{R} \phi$ piece. However, in the single-hole Eddington-Finkelstein 
metric, and hence in our conformal space, the behavior of $\hat{R}$ is 
such:  $\hat{R} > 0, ~ R \sim r^{-3}$ at infinity, with deviations at 
infinity from the Eddington-Finkelstein value that go as $r^{-4}$.  
Hence there is a finite contribution to the ``charge'' for the 
conformal factor equation arising from $R$, and $\phi$ differs only 
finitely from from the Eddington-Finkelstein value 
$\phi = 1$.

\section{Coordinate Conditions for Multiple Black Holes}

This section deals only with the case of boosted Eddington-Finkelstein 
(non-spinning) holes.  This development is prepared for immediate 
computational implementation of a multiple hole spacetime. As such, these ideas 
are tentative and need experimental (computational) verification. We present
two formulations of coordinate setting for multiple black hole systems.
Both are based on the idea that near the holes the spacetime should look as
closely as possible like a single hole Eddington-Finkelstein solution. In both
cases, we set boundary conditions or lapse and shift near the holes based on 
the single hole analysis above.

A defect of this current presentation is that some features of these 
coordinate specifications may be inappropriate for situations with net 
angular momentum (e.g. spiraling merger that settles to a Kerr black hole), 
but these coordinate specifications will suffice for evolution a finite 
time into the future, because they are \textit{most} accurate near the 
black holes, where the timescales associated with curvatures are 
shortest.

We first present a scheme which is very closely linked to the single hole
factor $H$.
First consider the single boosted hole case Eq.~(\ref{16}), with 
appropriate boundary conditions:  $H = 0.5$ at the horizon. $H$ 
satisfies a mixed boundary condition at infinity.  Specifying $H$ 
(solving Eq.( \ref{16}) for $H$) provides a complete solution for the 
gauge conditions, and guarantees that the algebraic relation between lapse 
and shift are consistent at the horizon with the Eddington-Finkelstein form.
Hence the horizon will be locked at a finite (moving, but not expanding or
contracting ) coordinate location.

Implementing this idea even in the single black hole case requires a 
kind of horizon-locking algorithm.  The logic is:

\begin{itemize}
	\item[--]  locate the coordinate position of the apparent horizon.

	\item[--]  solve Eq. (\ref{16}) for $H$, and use $H$ to determine 
	$\beta_a$ and $\alpha$. 
	
	\item[--]  adjust $\beta_a$ to move the horizon normally back to its 
	desired position.
\end{itemize}
Iteration over these steps may be required to correctly generate $H$ 
and $\beta_a$.

The multiple-hole case is complicated by the fact that ${\mathbf{v}}$ 
appears in Eq. (\ref{16}), and there is no uniquely defined ${\mathbf{v}}$ 
in a multiple-hole scenario.  We take the point of view that a) the 
velocity of a black hole against its surroundings is available by 
examining its recent history, b) for multiple hole systems the total 
momentum in the computational frame will be zero; where we compute the 
total momentum in a completely naive way as $\sum\limits_{i = 1, 
2}M_{i}{\mathbf{v}}_{(i)}$, where $M_{i}$ and ${\mathbf{v}}_{(i)}$ are the 
parameters used in the initial data setting.  

We are not claiming that this will give a final configuration which is on
average completely non-moving, but it seems likely to remove the majority of 
the residual average motion.  
Furthermore, computational science is experimental science, and the 
results of a run showing a final drift are a signal to adjust the initial 
total momentum-setting.  To apply coordinate ideas as in the preceding 
section, we would then expect that ${\mathbf{v}}$ in the coordinate 
specification Eq. (\ref{16}) is a function of position having a local value 
appropriate to black hole 1 when at the horizon of black hole 1; and smoothly 
changing to the value of ${\mathbf{v}}$ associated with hole 2.  A simple 
scheme would be to take the coordinate specification
\begin{equation}
	{\mathbf{v}} = {\mathbf{v}}_{1}\left(\frac{r_{h1}}{r_{1}}\right)^{p} 
		+ {\mathbf{v}}_{2}\left(\frac{r_{h2}}{r_{2}}\right)^{p}.
	\label{35}
\end{equation}
Thus velocities approach the appropriate velocity as the surface of 
the black hole is approached.  The quantities $r_{1}\,, r_{2}$ are 
computed as in Eq. (\ref{22}).  The quantities $r_{h1}\,, r_{h2}$ are 
the horizon radii along a ray to the point we wish to evaluate 
${\mathbf{v}}$.  The exponent $p$ will be chosen by experiment.  $p = 
1$ gives the situation simplest to analyze.  Then if we assume slowly 
moving black holes:
\begin{equation}
	{\mathbf{v}} = \frac{2M_{1}{\mathbf{v}}_{1}}{r_{1}} 
		+ \frac{2M_{2}{\mathbf{v}}_{2}}{r_{2}}\,.
	\label{36}
\end{equation}
By our ``zero momentum'' assumption, this is
\begin{equation}
	{\mathbf{v}} = 2M_1{\mathbf{v}}_{1}\left[\frac{1}{r_{1}} 
	- \frac{1}{r_{2}}\right].
	\label{37}
\end{equation}
For $r_{1} = r_{2}$, the point equidistant from the holes, one 
obtains ${\mathbf{v}} = 0$.  At infinity one has $v \rightarrow 0$ at 
least as $r^{-2}$.  At the surface of hole 1:
\begin{equation}
	{\mathbf{v}} = {\mathbf{v}}_{1}
	\left[1 - \frac{2M_{1}}{r_{2}}\right],
	\label{38}
\end{equation}
which gives a first order fractional correction to ${\mathbf{v}}$.  
This may require the use of a higher positive power $p$ in practical 
applications.

At intermediate locations we would expect $H$ to recognize the 
orbital angular momentum, and take on a Kerr-like form, but our 
present development supports only the Schwarzschild-like case.  Note, 
however, that the Kerr-like effects are ``mild'' until a final black 
hole forms, at which point the problem will become essential.

We mention here a second approach which 
can be considered: maximal slicing\cite{maximal}
(which determines the lapse $\alpha$ )
with minimal shear specification of the shift $\beta_i$.\cite{SmarrYork}  These 
elliptic equations require boundary conditions on lapse and shift.  
The boundary conditions specified in this section determine the horizon 
values of the lapse and shift, even when we solve them via methods other than
solving for $H$ as suggested above. Hence those boundary conditions
can be applied 
directly to the lapse and shift to be solved in the elliptic maximal 
slicing/minimal shear system.

\section{Discussion}

The description in section V of an algorithm to set binary black hole data
stars proceeds by generalizing the spatial metric to two centers, and 
erecting spatial ``unit'' vectors from those centers. With this form, a variant
of standard 3+1 data setting is used to set up the data. The exposition in
this paper concentrated on understanding the initial data structure of a 
single black hole, and describing how it can be used for a computational 
evolution. 
This method is to a large extent complementary to 
the other Alliance paper on setting data via the Kerr-Schild approach
\cite{bishop.1} assumes a general metric (even for two-hole data) of the form 
of Eq.~(\ref{1}), up to terms $O(t^2)$ where $t=0$ is the time of the initial
spatial slice. Furthermore, the initial value problem is solved in terms of
conditions on the null vector $l_\nu$, and on the multiplying scalar $H$,
rather than expressing it directly in terms of the usual 3+1 objects, 
$g_{ij}$ and $K_{ij}$. While the usual Kerr-Schild $l_\alpha$ is shear-free,
that condition is not imposed in this more general situation.
Because of the close connection of the approach in \cite{bishop.1} 
to the structure of the known Kerr-Schild solutions, a large number of 
cross-checks and analytical simplifications 
apply even in this more general situation. In part because if those
simplifications, \cite{bishop.1} is able to present data 
for perturbed Schwarzschild black holes.
An example set of data for "close" black holes is given in a perturbative 
limit where
the deviations from sphericity are small. In this case then analysis can be
carried through completely.

Attacking the binary black hole problem computationaly has sharpened our
perspective that all aspects of the relativistic problem must be understood,
for the computation to proceed. These two approaches, or some subsequent
combination, will be invaluable in setting and evolving binary black hole
data.
\section{Acknowledgment}

This work was supported by NSF grants \\ PHY9310083 and The Binary Black 
Hole Grand Challenge:  ASC/PHY9318152 (ARPA supplemented), by NSF metacenter 
grant MCA94P015P, and through computer time access through the Vice President 
for Research, The University of Texas at Austin.

\end{document}